\documentclass[utf8]{frontiersFPHY} 
\usepackage[dvipsnames]{xcolor}
\usepackage{url,hyperref,lineno,microtype,subcaption}
\usepackage[onehalfspacing]{setspace}

\usepackage[final]{changes}
\usepackage{comment}
\definechangesauthor[color=violet]{KPR}
\definechangesauthor[color=red]{red}

\def\keyFont{\fontsize{8}{11}\helveticabold }
\def\firstAuthorLast {Molnar~{et~al.}} 
\def\Authors{Momchil Molnar\,$^{1, 2, \dagger}$, Kevin P. Reardon\,$^{1, 2}$, Christopher Osborne\,$^{3}$, 
Ivan Mili\'c\,$^{1, 4}$}
\def\Address{$^{1}$ National Solar Observatory, Boulder, CO, 80303, Colorado, USA 
			
$^{2}$ Department of Astrophysical and Planetary Sciences, University of Colorado, Boulder, CO, 80303, USA
 
$^{3}$ SUPA School of Physics and Astronomy, University of Glasgow, UK

$^{4}$ Department of Physics, University of Colorado, Boulder, CO, 80303, USA

$^{\dagger}$ DKIST Ambassador
}
\def\corrAuthor{Momchil Molnar}
\def\corrEmail{momo@nso.edu}

\begin{document}
\onecolumn
\firstpage{1}

\title[Spectral deconvolution with deep learning]{Spectral
deconvolution with deep learning: removing the effects of spectral PSF broadening} 

\author[\firstAuthorLast ]{\Authors} 
\address{} 
\correspondance{} 

\extraAuth{}

\maketitle

\begin{abstract}

We explore novel methods of recovering the original spectral line profiles from data obtained by instruments that sample those profiles with an extended or multipeaked spectral transmission profile.
The techniques are tested on data obtained at high spatial resolution from the Fast Imaging Solar Spectrograph (FISS) grating spectrograph 
at the Big Bear Solar Observatory and from the Interferometric Bidimensional Spectrometer (IBIS) instrument 
at the  Dunn Solar Telescope. 
The method robustly deconvolves wide 
spectral transmission profiles for fields of view sampling a variety of solar 
structures (granulation, plage and pores) with a photometrical precision of less than 1\%. The results and fidelity of the method are tested on data 
from  IBIS  obtained using several
different spectral resolution modes. 

The method, based on convolutional neural networks (CNN), is extremely fast,
performing about
$10^5$ deconvolutions per second on a 
CPU and $10^6$ deconvolutions
per second on NVIDIA TITAN RTX GPU
for a spectrum with 40 wavelength samples. 
This 
approach is applicable for deconvolving large
amounts of data from
instruments with wide spectral transmission profiles, such as the Visible Tunable Filter (VTF) 
on the DKI Solar Telescope (DKIST). 
We also investigate its application to future instruments by recovering 
spectral line profiles obtained 
with a theoretical multi-peaked spectral transmission profile. 
 We further discuss the 
limitations of this deconvolutional 
approach through the analysis of the dimensionality 
of the original and multiplexed data.


\tiny
 \keyFont{ \section{Keywords:} Convolutional neural networks; Astronomical instrumentation; Spectroscopy;
 Deep Learning; Deconvolution algorithm.}
\end{abstract}

\section{Introduction}
\label{sec:Intro}

The finite spectral resolution of real instruments affects the
inferred signal by blending the intensities at different wavelengths. This
phenomenon is problematic for (solar) spectral lines, since the
shape of a line encodes essential information about a range of heights in the solar
atmosphere. 
However, some instruments use a lower
spectral resolution (broader spectral transmission profile) to increase instrument throughput and reduce integration times. 
Such a broad spectral point-spread function (sPSF) results fundamentally in a multiplexed sampling of the line profile, with the information from a given portion of the original spectral profile sampled multiple times at various positions in the sPSF (i.e. with varying relative attenuation) as the transmission function is tuned through the line.
This means that it should be possible to recover much of the underlying spectral information from this linear combination of samplings. 
The concept of exploiting this multiplexing to recover spectral information was originally developed by \citep{1979Caccin_reconstruction} and later \citep{1983Baranyi_profile_synth} in order to reconstruct spectral profiles sampled by the the relatively broad (0.15--0.5 \AA\ FWHM) sPSF of the tunable Universal Birefringent Filter (UBF)\citep{1975Becker_UBF}. The method developed, which relied on analytical descriptions of the sPSF, was employed by \citep{1983Caccin_imagingspec,1986Baranyi_Halpha} to reconstruct H-alpha and Na D line profiles recorded through a UBF. However, the data at the time were recorded on photographic film and the method was sensitive to noise and computationally demanding. 
The current observational demands for high-resolution imaging have resulted in instruments based on Fabry-P\'erot-interferometers that have sPSF's that are again suitable for this method.

\begin{figure}
    \centering
    \includegraphics[width=\textwidth]{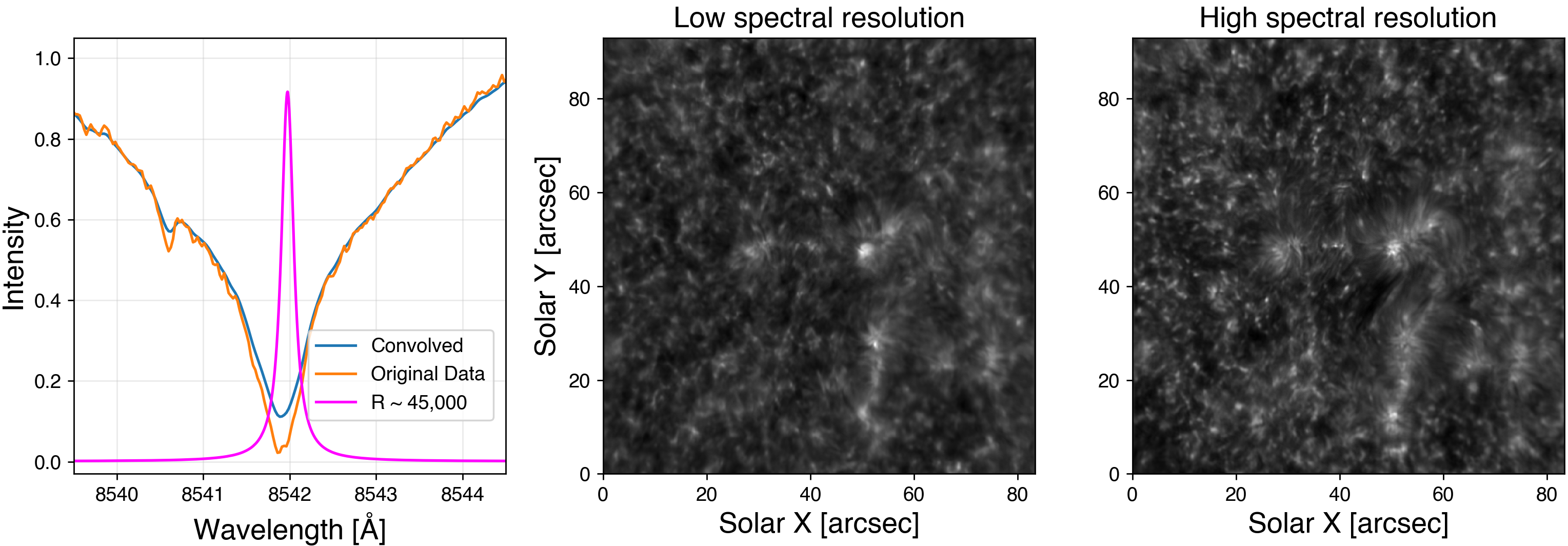}
    \caption{\emph{Left Panel:} Comparison of a sample
    line profile 
    of the Ca II 8542 \AA~line from the FISS dataset described in 
    Section~\ref{sec:deconvolve_SPSF}
    before 
    (\textcolor{orange}{orange}) and after (\textcolor{blue}{blue}) 
    convolution with the sPSF (\textcolor{magenta}{magenta}); The
    \textcolor{magenta}{magenta} curve is the transmission profile used 
    for convolving the orange profile to get the blue one 
    (corresponding to FP 2 of IBIS with R$\sim$45,000);
    \emph{Central and Right panels:} Comparison of chromospheric quiet
    Sun region
    observed with IBIS with low spectral resolution (R$\sim$50,000) on the central panel
    and with high spectral resolution (R$\sim$200,000) in the right
    panel.}
    \label{fig:intro}
\end{figure}

In this work, we
seek to evaluate machine-learning techniques that can retrieve higher-resolution spectra
from instrumentally broadened spectral profiles.
The effect of spectral smearing on the line shape
is shown in the left panel of Figure~\ref{fig:intro}
with an example spectrum of Ca II 8542 {\AA} from the 
FISS/BBSO\cite{2013SoPh..288....1C} spectrograph.
The orange line is the spectrum as observed with the full
spectral resolution of FISS of about 
R$\sim$150,000 (where spectral resolution R is defined as the wavelength of observation divided by the FWHM of the profile). 
Instead, the blue line shows the same spectrum convolved with
a Lorentzian-shaped sPSF with R$\sim$45,000. Given the typical shape of an absorption line, the convolution with a broad sPSF
raises the intensity around the line core and broadens the wings of the profile. 

This smearing tends to increase the similarity among
different spectral
profiles, also reducing the spatial contrast 
and the ability to identify small scale
structures in images of the solar atmosphere. An example
of this is presented in Figure~\ref{fig:intro}
(central and right panels), 
with observations in the core of the Ca II 8542 {\AA} 
line from the Interferometric Bidimensional Spectrometer (IBIS)~\cite{2006SoPh..236..415C}
instrument at the Dunn Solar Telescope 
The same 
FOV was observed at the instrument's normal high spectral resolution (R$\sim$200,000), but also at a much lower spectral resolution (R$\sim$50,000), 
which was achieved by removing the ``narrow passband'' Fabry-P\'erot 
interferometer (FPI) from the optical path \cite{2008A&A...481..897R}.
We can see the reduced contrast in the FOV with lower spectral resolution 
which deteriorates the identification of the chromospheric features.
Hence, mitigation of the degraded spectral purity of our
observations is essential for furthering our understanding of the Sun.

Furthermore, the compressible nature of 
spectral lines as suggested by 
\cite{Asensio_Ramos_2007} could 
allow the sampling and subsequent recovery of the
full spectral profiles with a lesser number of measurements
by using a suitably adopted measurement basis. This approach
could improve instrumental performance by increasing the
sampling cadence through a reduction in the 
number of instrumental tuning steps needed to sample the line.

In this paper we perform experiments to test the applicability of 
Convolutional Neural Networks (CNNs)
to perform the de-multiplexing of spectral line profiles in different 
scenarios. We examine the photometric
accuracy than can be achieved with these techniques.
Finally, we discuss the limitations on the precision
of the recovered profiles based on the 
dimensionality of the data derived from 
maximum-likelihood
intrinsic-dimensionality estimate~\citep{Levina_Bickel_2005}. 

\section{A deep learning approach}
\label{sec:deep_learning}

We utilize deep Convolutional Neural Networks (CNNs) for the 
deconvolution process as they are
powerful function approximators which are widely used for pattern recognition
and image processing\cite{Goodfellow-et-al-2016}. We
used an encoder-decoder architecture because it
can extract the relevant features from noisy data
(encoder) and then recreate the signal of interest
from the latent space (decoder). 
The architecture of the network consists
of three convolutional layers, 
followed by three symmetric upsampling 
(``deconvolutional'') layers, followed by two
dense layers with dimensions of
the output data. The three consecutive 
convolutional layers (and their corresponding
upscaling layers) have 
[5, 10, 20] filters and used a three-pixel kernel.
Rectified Linear unit (RELU) activation function was used
for all layers with the exception of the last one
where we have used a linear activation function\cite{he2016deep}.
Furthermore, in Section~\ref{sec:deconvolve_SPSF} we add the 
input layer to the last dense layer of the network to improve 
the performance of the network.
This is due to the
fact that in this architecture, the network has to
estimate only the corrections 
to the convoluted signal instead of recreating 
the whole spectral line profile.
However, this will cause the core of the spectral profile to
be poorly fit since the most significant corrections
are needed there (as can be seen in left panel of Figure~\ref{fig:intro}).
To alleviate this issue, we
introduce a custom loss function which is a weighted mean square error. 
The weights are
chosen inversely proportional to the intensity of the line profile so as
to emphasize the precision of the recovered profiles in the line core.
We trained our network with the 
Adam optimizer~\cite{kingma2014adam} for about one thousand epochs before
satisfactory convergence was achieved. The network
was implemented with Keras under
Tensorflow~\cite{tensorflow2015-whitepaper} and can be found in the public
repository of the project.

\section{Spectral deconvolution with CNN}
\label{sec:deconvolve_SPSF}

\subsection{Deconvolution of synthetic data from FISS}
\label{subsec:deconvolve_FISS}

\begin{figure}
    \centering
    \includegraphics[width=\textwidth]{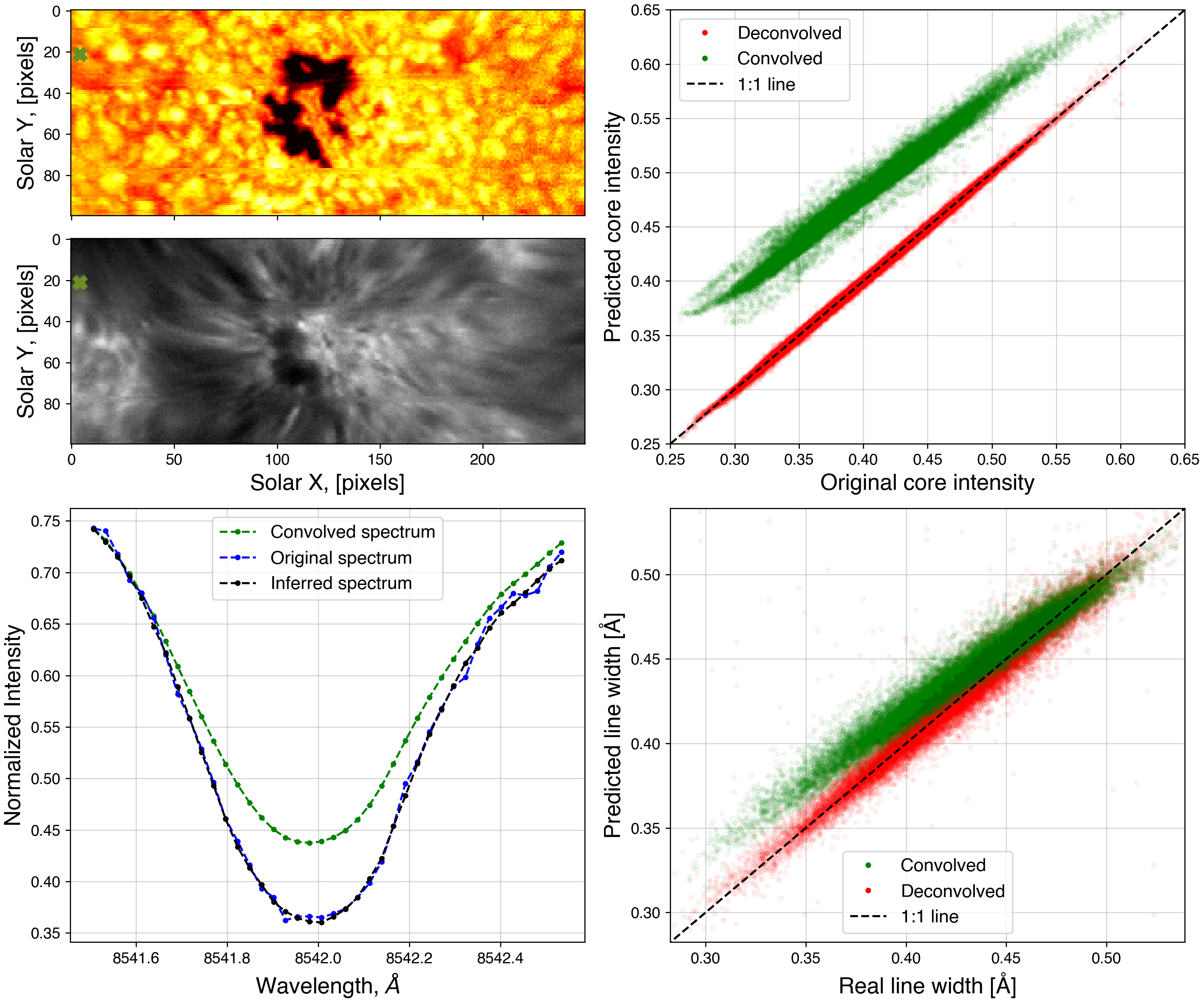}
    \caption{\emph{Top left:}
    Continuum image/ Line core intensity of Ca II 8542 {\AA}
    of the FISS dataset used for the experiment
    in Sections \ref{subsec:deconvolve_FISS} and \ref{sec:recover_multiplex_profiles}; 
    \emph{Bottom left:} A sample profile (coordinates
    [4, 21] in our datacube, green cross in top left panel) shown before convolution with
    wide SPSF (\textcolor{blue}{blue line}), 
    after the convolution (\textcolor{green}{green})
    and after the  deconvolution with the CNN (black); 
    \emph{Top right:} Comparison of the line core intensity
    recovered with the algorithm -- 
    true line profile (black line is the one-to-one line). 
    The approach for measuring line intensity and width are
    described in section~\ref{subsec:deconvolve_FISS}. 
    \emph{Bottom right:} Same as the previous panel but for the 
    line core width of the Ca II 8542 \AA~line.}
    \label{fig:SPSF-deconvolve-FISS}
\end{figure}

To test the CNN approach for sPSF deconvolution, we utilized Ca II 8542 \AA~
data from the FISS/BBSO~\cite{2013SoPh..288....1C}
instrument (R$\sim$150,000) obtained on June 22, 2016. 
We created a training set by 
convolving each spectral profile with a Lorentzian sPSF with an effective 
R $\sim$ 45,000 (corresponding roughly to the FWHM of IBIS's FPI \#2 transmission profile\cite{2008A&A...481..897R}). 
The bottom left panel of 
Figure~\ref{fig:SPSF-deconvolve-FISS} shows a sample 
profile from the FISS
instrument in blue and convolved with the FPI \#2 profile in green.
The CNN was trained with spectra from a single raster scan 100 
$\times$ 250 spatial pixels corresponding to 16 $\times$ 20 arcseconds  on the Sun centered on a pore near disk center which took 16.5 seconds.
Satisfactory convergence was accomplished in about one thousand epochs
with the relative RMS error at the last epoch of the training reaching about $1.5 \times 10^{-4}$. 

The performance of the CNN was
tested on a different raster (data not seen 
by the network previously) from FISS of the same region of the Sun acquired \textit{5} minutes after the scan used for training. 
The line core intensity value and position was determined as the minimum of a
parabola fitted to the 7 points around the pixel position with the lowest intensity. 
The core width of the line profile was measured 
(following~\cite{2009A&A...503..577C}) as
the bisector width at the intensity equal to half 
the difference between 
intensity of the line core and the intensity at 
a fixed offset of 0.4 {\AA} from the wavelength position of the line core.

The algorithm achieved 0.76 \% precision photometry of the line core 
intensity and 
2.5 \% precision retrieving the line core width.
These results are illustrated in the right column of
Figure~\ref{fig:SPSF-deconvolve-IBIS}. This example shows the 
robustness of the ML approach for retrieving spectral line profiles.
The algorithm takes about $7 \times 10^{-6}$ seconds for a single
inversion of 40 point spectrum on an Intel i7-4780HQ CPU and only
$0.3 \times 10^{-6}$ seconds on a NVIDIA TITAN RTX
GPU. We take 
into account the I/O overhead for the GPU
inversions, as we used
a dataset of 16 million spectra with
40 wavelengths points (similar to the VTF full CCD readout) which amounts to about 20\% of
the memory of the GPU. This 
method is slightly faster than
the scipy.signal.deconvolve algorithm which uses 
a digital filter, but the latter 
cannot reproduce the wings of the line well due to
boundary effects. Compared to more computationally intensive algorithms 
such as the 
Richardson-Lucy~\cite{Richardson:72} deconvolution algorithm, we found 
that our algorithm is about 100 times faster. Furthermore,
it does not require fine tuning of parameters once a suitable training set is
provided.

\subsection{Deconvolution of real spectral data from IBIS}
\label{subsec:deconvolve_IBIS}

\begin{figure}
    \centering
    \includegraphics[width=1\textwidth]{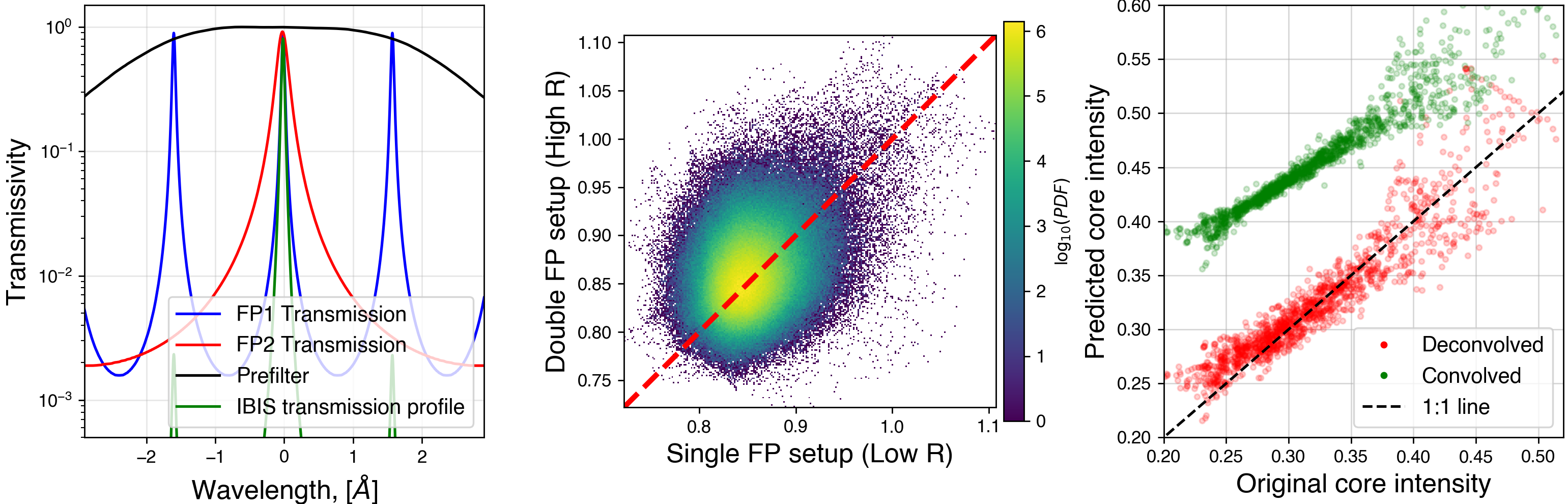}
    \caption{\emph{Left panel}: Transmission profile of the components of the IBIS Instrument centered around 
    8542 {\AA}. The transmission profiles of the two Fabry-P\'erot etalons are in \textcolor{red}{red} 
    and \textcolor{blue}{blue}; the 8542 {\AA} prefilter profile is the black line; The effective transmission 
    profile of the instrument is presented in \textcolor{olive}{green};
    \emph{Central panel}: Histogram of 
    the wing intensity in the data sets with single FP (low R) 
    and both FPs (high R) in the optical system;
    \emph{Right panel}: The result of the deconvolution algorithm applied 
    to real IBIS data (same as top right panel in Figure~\ref{fig:SPSF-deconvolve-FISS})
    for the line core intensity.}
    \label{fig:SPSF-deconvolve-IBIS}
\end{figure}

To test the method on real Fabry-P\'erot data we obtained a dataset with 
the IBIS instrument at the DST with high (R $\sim$ 200,000) and low
(R $\sim$ 50,000, similar to the the FISS tests above) spectral resolution of
the same region of the Sun. We achieved the 
different spectral resolutions by 
utilizing the fact that the IBIS instrument consists of two 
Fabry-P\'erot (FP) interferometers in series, one of which has a profile three times narrower than the other (the components of the IBIS instrument are presented in the left panel of
Figure~\ref{fig:SPSF-deconvolve-IBIS}). Hence,
if we take the narrower FP (FP \#1) 
out of the optical path,
we obtain observations with a lower spectral resolution.
We imaged a region near disk center of the Sun in the Ca II 8542 {\AA} line,
where we scanned a spectral region of 4.4 {\AA} centered around the 
line core with a spacing
of 50 m{\AA}. We acquired five separate exposures at each 
wavelength (for post-processing MOMFBD~\cite{MOMFBD} reconstruction to minimize seeing effects) which resulted in two
datasets of the same solar structures with different spectral resolution taken 4 minutes apart. 
We applied the deconvolution algorithm to the these datasets using as the input
the lower spectral resolution data obtained with a single FP and as the expected output the higher spectral
resolution dataset obtained with both FPs. Images from the two datasets are presented in 
the central and right 
panels of Figure~\ref{fig:intro}.

We had limited success with deconvolving this dataset  
as the spectral profiles had changed significantly even over the 4 minute interval 
between the datasets. To illustrate this, the central panel of
Figure~\ref{fig:SPSF-deconvolve-IBIS} shows the density plot of the 
quasi-continuum in the wings of the two datasets. 
The lack of obvious correlation (confirmed by
visual inspection of the data)
 shows that the structures on 
the solar surface have significantly changed between 
the two datasets were obtained (consistent with granular lifetimes of $\sim$8 minutes).

To explore the validity of this deconvolution 
approach, we chose a
subset of spatial pixels from the IBIS
scans based on a criteria to identify spectral profiles that
did not change significantly between the two samplings.
This step allows the CNN to train primarily on the effects from the spectral smearing, 
not the evolution of the solar
atmosphere. The imposed criteria are that the measured Doppler 
velocity change\footnote{For symmetric line profiles, 
Doppler velocity does not depend on R.}
between the two consecutive 
samplings is no greater than half a resolution element (0.6 km/s)
and that 
the location of the pixel in the cumulative distribution of 
intensity and width (relative to the other pixels sampled at the same spectral resolution) does not change by more than 5 percentiles. 
The expected versus deconvolved core intensities are presented in the right
panel of Figure~\ref{fig:SPSF-deconvolve-IBIS} 
(compare to 
Figure~\ref{fig:SPSF-deconvolve-FISS}). The scatter is larger than the FISS synthetic data due to the effects of solar surface evolution occurring between the acquisition of the two components of the training set, whose correction is 
beyond the scope of this project.
Future tests of this method could 
emphasize obtaining a more nearly cotemporal training dataset  
by restricting the range of the spectral 
scan around line core, reducing the time separation between the 
different spectral resolution scans significantly. We note that distinct 
training datasets, derived at different times or even using a separate 
instrument (e.g. a slit spectrograph), could be applied to multiple 
datasets obtained with a low-spectral-resolution FP instrument 
(e.g. VTF~\cite{2012SPIE.8446E..77K} at the 
DKIST~\cite{2016AN....337.1064T}), under the assumption that: 
a) the sPSF and straylight of each instrument is well characterized; 
and b) the same general types of solar spectral profiles are sampled 
in both cases.

\section{Recovering undersampled spectral profiles with multi-peaked sPSF}
\label{sec:recover_multiplex_profiles}

It was suggested by
\cite{Asensio_Ramos_2007} 
that not all 
wavelength points in a spectral line are linearly
independent and 
that recovery of a full spectral profile with lesser number of 
measurements (in a suitably chosen basis) is possible. This presents us with the 
opportunity to extract useful information from undersampled spectral line 
profiles which can result in more efficient spectral sampling or better
compression techniques for space-based missions.

\begin{figure}
    \centering
    \includegraphics[width=\textwidth]{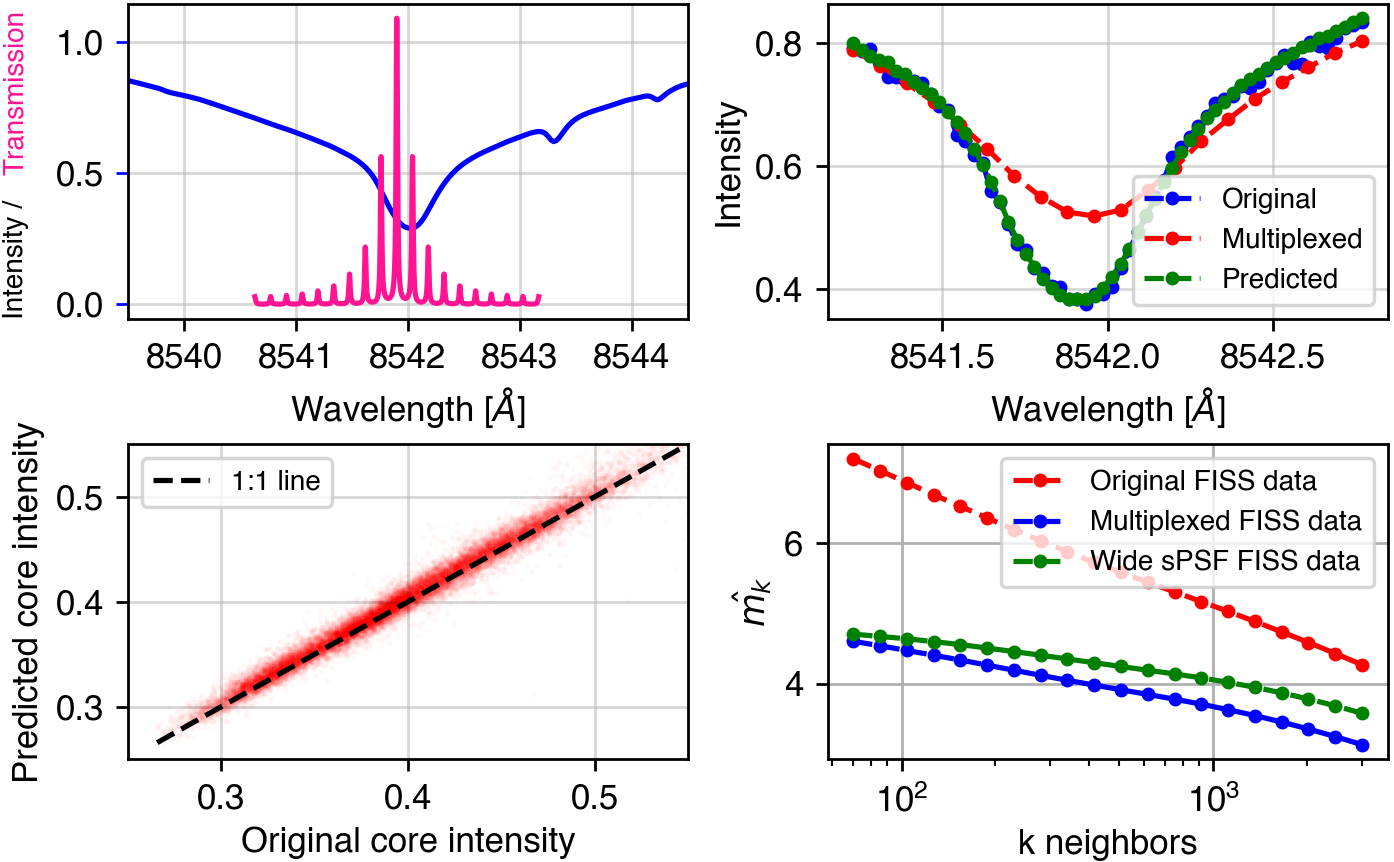}
    \caption{\emph{Top left}: The spectral transmission profile is
    (\textcolor{magenta}{magenta} scaled by 9 for best representation) applied 
     to the spectral profiles overlaid over the 
     average Ca II 8542~{\AA} line.
     \emph{Top right}: Sample spectral profile from the multiplexing (\textcolor{red}{red}),
     the retrieved spectral profile (\textcolor{green}{green}) and the original spectral 
     profile (\textcolor{blue}{blue});
    \emph{Bottom left}: Retrieved line core intensity from this approach vs the original 
    line core intensity; 
    \emph{Bottom right}: Maximum Likelihood Intrinsic
    Dimensionality Estimate for the original FISS data, the multiplexed
    FISS data and the data convolved with the wide sPSF in
    Section~\ref{subsec:deconvolve_FISS}.}
    \label{fig:multiplexing}
\end{figure}

To test the ability of CNNs to recover spectral profiles
multiplexed with multipeaked transmission profiles as suggested 
in~\cite{2010A&A...509A..49A}
we created a transmission profile of a hypothetical dual
Fabry-P\'erot interferometer with spacings of the
etalons of 2.6 and 0.058 centimeters with 0.99 reflectance coatings. 
The resulting transmission profile of this hypothetical 
instrument
is presented in the top left 
panel of Figure~\ref{fig:multiplexing} overplotted on the 
Ca II average line profile. The transmission profile was designed such
that the higher-spectral-resolution FP generates multiple peaks 
within the chromospheric core of the solar spectral line while the 
lower-spectral-resolution FP selects a limited 
range such that 80\% 
of the transmitted light is coming from three central peaks.
The properties of the FPs were chosen 
to optimize the precision of the deconvolutions. 
If the peaks of the transmission profile are too close or too far apart,
the neural network's performance drops. Further
optimization of 
the FP setup can be achieved through exploration of the
dimensionality of the data as described in the following
paragraphs.

We applied the transmission profile to the 
FISS data used in Section~\ref{subsec:deconvolve_FISS}
where we downsampled the number of spectral samples by 3 for this
particular example.
A sample deconvolution is presented in the top right panel of 
Figure~\ref{fig:multiplexing}, which shows a good agreement between
the original and deconvolved spectral profiles.
The bottom left of Figure~\ref{fig:multiplexing} 
shows the scatter of the derived line core intensity of the 
multiplexed line versus the original line core intensity.
We achieve a RMS
of the retrieved 
line core intensity of about 2\% for this numerical setup.
This is about three times worse than the previous experiment with 
FISS data in Section~\ref{subsec:deconvolve_FISS}. 
Our result is close to the precision obtained by
Asensio Ramos (2010)~\cite{2010AN....331..652A} where the author uses
a single FP etalon with a prefilter.

To explain the lower precision of this multipeaked-multiplexing deconvolution 
approach compared to the deconvolution of the wide sPSF in 
Section~\ref{sec:deconvolve_SPSF}, we evaluate
the dimensionality of the data. 
The dimensionality quantifies how much information is
contained in the observations and can be used to evaluate the 
losses due to the spectral multiplexing.
We computed the
maximum likelihood intrinsic estimated dimensionality (MLIED, introduced by~\cite{Levina_Bickel_2005} and suggested
for spectroscopic use by~\cite{Asensio_Ramos_2007}), which is an 
estimate of the dimensionality of the data based on phase density distribution.
The bottom right panel of 
Figure~\ref{fig:multiplexing} shows the dimensionality 
estimate for the original data,
the multiplexed data, and the data convolved with a wide sPSF
versus the number of neighboring spectra used for the computation of the dimensionality.
We find that the dimensionality of the multipeaked-multiplexed data is lower than
the data convolved with a wide sPSF, while 
the original data has the highest dimensionality. 
It is expected since 
the convolution process introduces a loss of information.
This greater loss of information is why the multipeak approach 
(as modeled in this section) results in a
lower precision compared to the results 
those for for a single, broad sPSF.

This approach, evaluating spectral dimensionality, could be used in future design studies of instruments
 as a way toward building more efficient instruments, optimizing throughput and preservation of spectral information.
Further work is needed to connect the dimensionality analysis (and resulting choice of instrumental sPSF) with the accuracy and precision of the retrieval of physical information from the spectral profiles via the optimal choice of parameters for
the FP system.

\section{Conclusions and future work}
\label{sec:conclusions}

We have presented a novel way to perform deconvolution of 
spectral data with deep
learning. Our method is robust and reliable if the 
sPSF of the instrument
is well known \emph{a priori} and we have a
reliable training set. Our method can deconvolve a single, 
40-wavelength spectrum in 0.3  
microsecond on a NVIDIA TITAN RTX GPU with a photometric 
precision of the line core intensity of less than 1$\%$.
The speed of the proposed algorithm
makes it very effective for processing large 
numbers of spectra, with further improvements possible 
if the deconvolution is performed on batches of
data on a GPU. 
With the next generation of solar instruments (such as the
VTF at the DKIST),
which will produce terabytes of spectral data per day, the speed of
deconvolutional techniques will become increasingly important.

The technique was demonstrated here only for non-polarized spectroscopic measurements, but full spectropolarimetric measurements (including also the spectral dependence of the circularly and linearly polarized components of the signal) are a key aspect of observational solar science. There is no conceptual reason why this method could not be extended to the measurement of the Stokes profiles, given suitable training sets. However, since the polarized components of the signal tend to be just a small fraction of the overall signal (a few percent or less), any systematic errors introduced into the deconvolved profiles might bias the recovery of the information about the magnetic field. Future work will evaluate the application of this method to this common usage scenario.

We successfully recovered spectral line profiles observed with a
multipeaked spectral transmission profile, as suggested before 
in~\cite{2010A&A...509A..49A}, using 
a theoretical dual Fabry-P\'erot etalon instrument with optically
realistic parameters.
Our numerical experiments showed that
a careful choice in the separation of the 
peaks of the transmission profile
allows the retrieval of the spectral line
profiles with a photometric precision of about $\sim 2 \%$
while requiring 3 times fewer spectral samples. 
This could be used in the design of future Fabry-P\'erot based
instruments that would require fewer measurements (higher
cadence) and potentially have higher transmission 
(shorter exposure times).

Future work will include
obtaining a more suitable dataset for improving the results from 
the experiment with IBIS data in Section~\ref{subsec:deconvolve_IBIS}. 
In order to apply
this deconvolutional approach to real observations in a routine manner 
we will need training
sets consisting of low and high resolution data of a variety of 
regions on the Sun. One approach
would be to obtain 
nearly simultaneous observations of the same region of the Sun with 
low and high 
spectral resolution instruments at comparable spatial resolution.
Another feasible way to create the training dataset is by numerically 
convolving data from a high-spectral-resolution instrument with 
the known sPSF of the low-spectral-resolution instrument to 
generate simulated observations.
Both approaches have advantages and disadvantages 
but provide alternative approaches to real world
applications of this method.
We therefore hope that future 
instruments will consider the approaches described here and 
in~\cite{2010A&A...509A..49A} to leverage the
advantages of machine learning and compressive sensing to more 
efficiently retrieve information from the solar spectrum and further our
understanding of the Sun.
	
\section*{Conflict of Interest Statement}

The authors declare that the research was conducted in the absence of any commercial or financial relationships that could be construed as a potential conflict of interest.

\section*{Author Contributions}

KR proposed the setup of the two problems and suggested the approach to solving
it.
KR obtained the FISS data. KR and MM acquired the IBIS data at the DST. MM reduced the
IBIS
data and constructed the CNNs with help from CO and IM. MM performed the
tests of the accuracy of the methods. 
All authors contributed to the manuscript.

\section*{Funding}
MM was supported for the work on 
this article from the GEH fellowship provided from the University of Colorado, Boulder.
Funding for the DKIST Ambassadors program is provided by the
National Solar Observatory, a facility of the National Science
Foundation, operated under Cooperative Support Agreement
number AST-1400405.
CO acknowledges support from the UK’s Science and Technology Facilities Council (STFC) doctoral training grant ST/R504750/1.

\section*{Acknowledgments}
MM would like to thank the organizers of the ``ML in Helio'' 
conference for the support to attend this great workshop. 
CO is grateful to the members of the National Solar Observatory for many scintillating discussions.
The authors would like to thank Kyeore Lee, Jongchul Chae, and 	Kwangsu Ahn for generously providing the FISS spectra.
Furthermore, the authors would like to thank the referees for their comments which helped improve the manuscript and Andr\'es Asensio Ramos
for the useful discussions 
which helped with the CNN approach and
the Maximum Likelihood Dimensionality Estimate. The National Solar Observatory (NSO) is operated by the Association of Universities for Research in Astronomy, Inc. (AURA), under cooperative agreement with the NSF. IBIS has been designed and constructed by the INAF/Osservatorio Astrofisico di Arcetri with contributions from the Universit\`a di Firenze, the Universit\`a di Roma Tor Vergata, and upgraded with further contributions from NSO and Queens University Belfast.

\section*{Data Availability Statement}
The datasets and the code for this study can be found in the github repository of the author \url{https://github.com/momomolnar/SPSF_remove} . 

\bibliographystyle{frontiersinHLTH&FPHY} 
\bibliography{test}

\end{document}